# Decision Algorithm for Data Center Orbital Angular Momentum Receiver

Judy Kupferman and Shlomi Arnon
Electrical and Computer Engineering Department
Ben-Gurion University of the Negev
P.O Box 653 IL-84105 Beer-Sheva ISRAEL
E-mail: shlomi@bgu.ac.il

*Abstract* — We present a new scheme for an OAM communications system which exploits the radial component *p* of Laguerre Gauss modes in addition to the azimuthal component *l* generally used. We derive a new encoding algorithm which makes use of the spatial distribution of intensity to create an alphabet dictionary for communication. We suggest an application of the scheme as part of an optical wireless link for intra data center communication. We investigate the probability of error in decoding, for several detector options.

## I. INTRODUCTION

Light beams were shown to have orbital angular momentum (OAM) in 1992 [1]. The OAM of light can be decomposed into a set of orthonormal modes, making them an exciting new tool for communication capable of meeting the growing demand for higher bandwidth. One such set of modes are the Laguerre Gauss (LG) modes, characterized by both azimuthal and radial parameters. Use of OAM in communications schemes generally exploits the azimuthal parameter. Here we propose a scheme for OAM communications using both parameters of LG modes, and we suggest a system for use in data centers where it could be especially useful.

Use of OAM for communication systems has been extensively studied [2,3,4,5,6,7], for applications in various situations including quantum information, optical tweezers, and communications under water [3,4, 8,9,10]. A comprehensive review appears in [2] outlining the application of OAM to communications: beam generation, multiplexing, transmission through free space and fiber. A hybrid fiber/free-space communications system was described in [11].

A new application is here proposed for use of OAM in data centers. Data centers are at the basis of the technology behind the information revolution, and spearhead the need for larger bandwidth. In recent years, optical wireless links have been proposed as a way to improve the flexibility of the network as part of an optical wireless and fiber hybrid system. We suggest that orbital angular momentum can provide a novel practical solution for such a system [12,13,14,15,16].

OAM optical wireless communication (OWC) can fulfill a number of requirements in DC. First, DCs need to be scaled up, and it is desirable to have a flexible network that can be reconfigured for change in load and traffic. This can be done with using transceivers on top of racks as part of an optical wireless and fiber hybrid communication network [12,13,14]. A second consideration is energy consumption: DCs consume a large amount of energy, but on average, one in three DC servers is nonoperational at any given time [15]. If network topology could change quickly, energy consumption could be reduced by rerouting workloads from partially loaded servers, and thus enabling idle servers to shut down. A third consideration is bandwidth; since OAM can be produced in a large number of modes, there is a possibility for a large alphabet for encoding. In addition, an OAM OWC system would require almost no maintenance. In intra-DC communications distances are short, and the environment is controlled [17], and an OAM OWC system is a very promising solution for the growing demands of DC communication systems.

OAM light beams have a helical phase pattern and can be prepared in modes that are mutually orthonormal. Generally the azimuthal (phase) component of such modes has been used for communication. Here we present a system for Laguerre Gauss OAM modes, which exploits the spatial intensity distribution, which is a function of both azimuthal and radial mode components. Since distances in DCs are relatively short, the receiver system covers the entire beam. Thus, these modes can provide a very large alphabet for encoding information, enabling an increase in channel capacity of communication systems, and the beams can be efficiently multiplexed and demultiplexed. In [18] we proposed a system for an OAM OWC link in a DC which was based on local threshold detection and decision making. Here we suggest an alternative method of detection, where the information is decoded by means of the algorithm we propose below.

In order to process information transmitted with OAM light, it is necessary first to detect which modes have been transmitted, out of a large set of modes, and then to decode the information.. A specialized device for sorting and decoding OAM modes has been designed [19,20]. Our proposal is for a more modest implementation which does not require specialized mode sorting equipment. In this method each transmitted OAM beam reaches a set of four detectors, which were shown in [18] to be sufficient for the system in question. Signals from the detectors are compared to a prepared dictionary of modes.

This paper is organized as follows. Section II contains a summary of the basic theory of OAM. In Section III we describe a scheme for OAM communication and the algorithm for analysis of the system, and conclusions follow in Section IV.

## II. ORBITAL ANGULAR MOMENTUM

Allen et al [1] showed that beams with an azimuthal phase term $\exp[il\phi]$ have an OAM of $l\hbar$ per photon (where $l$ is topological charge, $\phi$ is the azimuthal angle, and $\hbar$ is Planck's constant $h$ divided by $2\pi$). OAM can be decomposed into a complete set of orthonormal Laguerre-Gauss modes which solve the paraxial Helmholtz equation in cylindrical coordinates. These have amplitude $A_{klp}(\vec{r})$ and phase $\Theta_{klp}(\vec{r})$ as follows [21]:

$$A_{klp}(\vec{r}) = \frac{N_{lp}}{w_0\sqrt{1+\frac{z^2}{z_R^2}}}\left(\frac{\sqrt{2}r}{w(z)}\right)^{|l|} L_p^{|l|}\left(\frac{2r^2}{w(z)^2}\right) e^{-r^2/w(z)^2} \quad (1)$$

$$\Theta_{klp}(\vec{r}) = \frac{kr^2 z}{2(z^2+z_R^2)} + l\phi + (2p+|l|+1)\arctan\left(\frac{z}{z_R}\right) + kz$$

where $z_R$ is the Rayleigh length, $w(z)$ is the beam width at axial coordinate $z$, and $w^2(z) = 2(z^2+z_R^2)/kz_R$, $N_{lp} = \sqrt{2p!/\pi(|l|+p)!}$ is a normalization constant, $L_p^{|l|}$ is the associated Laguerre polynomial and $A_{k00}$ is the amplitude for a plane wave propagating along the $z$ axis with wave vector $k$.

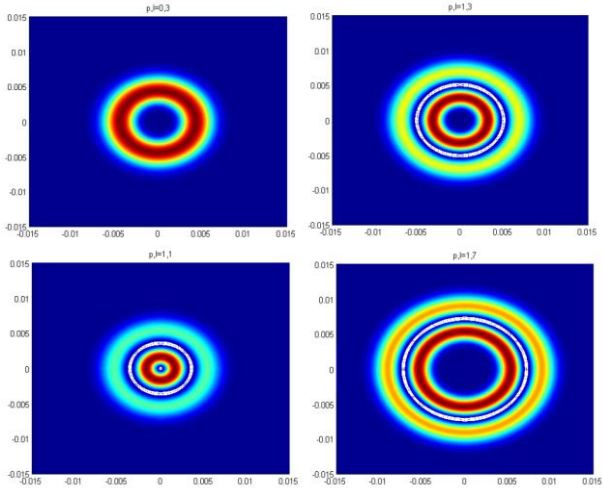

Fig. 1. Beam patterns for four different Laguerre-Gauss modes. Top, left to right: P=0, L=3 and P=1, L=3. Bottom, left to right: P=1, L=1 and P=1, L=7.

The azimuthal and radial indices, l and p, appear in both amplitude and phase. Direct detection of an OAM signal does not preserve phase information, but the amplitude alone is sufficient to characterize different modes for purposes of communication. Figure 1 shows the beam patterns for four different Laguerre-Gauss modes. OAM modes can be generated in a variety of ways including holograms, cylindrical lens converters, phase plates, spatial light modulators (SLMs) and digital mirror devices (DMDs) [11]. In this paper we deal specifically with Laguerre-Gauss (LG) modes but the concept may be extended to other mode families.

## III. RECEIVER DECISION ALGORITHM

In this paper we present a scheme for a ring detector receiver (Figure 2). The detector is composed of PIN photodiodes arranged in a series of annular rings that fill the entire beam area. Here we propose a performance decision algorithm based on minimum distance between the signal detected from the incoming beam, and one of a set of LG modes. The input beam is detected as a set of four signals. This is matched against a previously prepared set of arrays, which constitute a numerical dictionary of LG modes, to determine which mode has been transmitted and to extract the information. The system is arranged as follows: At the transmitter, laser light is converted to an LG mode according to the information input. The beam is then collimated by a telescope, and transmitted for a short distance in free space to a telescope at the receiver unit. There the transmitted beam is directed to an array of annular detectors. The design for the proposed receiver takes advantage of the symmetry of LG beam, and so the detector has radial symmetry. (See Figure 2).

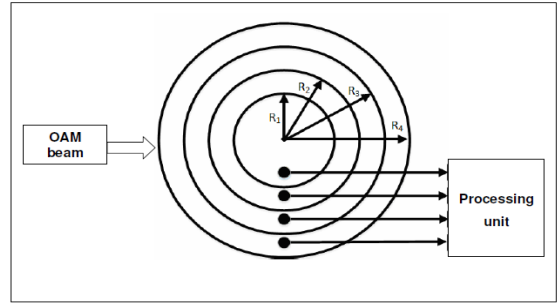

Fig. 2. The OAM receiver scheme. The detector is a set of concentric rings filling the entire beam area.

For this proposal we chose four modes (pictured in Figure 1) which can be easily differentiated at a relatively small radius. The detector ring radii are evenly spaced so that all the transmitted power lies within the rings. Each detector signal is an integral over the normalized beam intensity of that ring. Figure 3 is a graph of intensity distribution for an alphabet of the four LG modes from Figure 1. The graph shows the different intensity peaks at various distances from the center. The four detector signals are compared to a previously prepared dictionary of vectors each representing a different LG mode, and in this way the correct mode is identified and the information can be decoded.

The detector signal is analyzed as follows. Signals from the four detector rings are collected as a vector $\vec{M}$, whose components are

$$M_1 = \int_0^{R_1} 2\pi r \left|A_{klp}(r)\right|^2 dr$$

$$M_{i=2,3,4} = \int_{R_{i-1}}^{R_i} 2\pi r \left|A_{klp}(r)\right|^2 dr, \quad (2)$$

where $A_{klp}$ is the field amplitude that appears in equation (2). The dictionary of the four LG modes consists of four vectors $\vec{LG}(pl)$ according to each of the four LG modes in question, shown in Figures 1 and 3. The components of each $\vec{LG}(pl)$ vector are the values numerically found for an integral of that LG mode over the radius of each detector ring, corresponding to the components $M_1$, $M_2$, $M_3$, $M_4$ from eq. (2). Table 1 shows normalized intensity for each ring of the detector,

| Mode | Ring 1 $M_1$ | Ring 2 $M_2$ | Ring 3 $M_3$ | Ring 4 $M_4$ |
|---|---|---|---|---|
| P=1,L=1 | 0.31956 | 0.47743 | 0.20212 | 0.00088 |
| P=0,L=3 | 0.07653 | 0.79962 | 0.12346 | 0.00039 |
| P=1,L=3 | 0.152689 | 0.32658 | 0.50783 | 0.01290 |
| P=1,L=7 | 0.00133 | 0.37773 | 0.39541 | 0.22474 |

Table 1: Intensities of detector rings for four modes

The intensity was computed by squaring the beam amplitude and integrating over ring area. Thus the intensity for each mode is a vector with four components, one for each detector ring. In order to determine which is the correct input mode, the distance D is calculated between the input vector $\vec{M}$ and each of the four mode vectors:

$$D = |\vec{M} - \vec{LG}(pl)| = \sqrt{\sum_i (M_i - LG(pl)_i)^2} \quad (3)$$

The LG mode with minimal distance from the input vector is chosen as the correct one.

We calculated performance of the system with a noisy input signal. Since hot aisle distances in DC's range from about 1-2 meters [6], propagation might be to a receiver across the aisle, or several racks down. In a numerical simulation we took a 2m propagation distance and Rayleigh length of 2.03m. The wavelength was set at 1550 nm for favorable transmission in free space [2], and beam waist was 1mm ($\lambda z_R / \pi = w_0^2$). We constructed a dictionary of the four modes shown in Figures 1 and 3. This consisted of four arrays whose elements were the normalized intensities integrated over each detector ring.

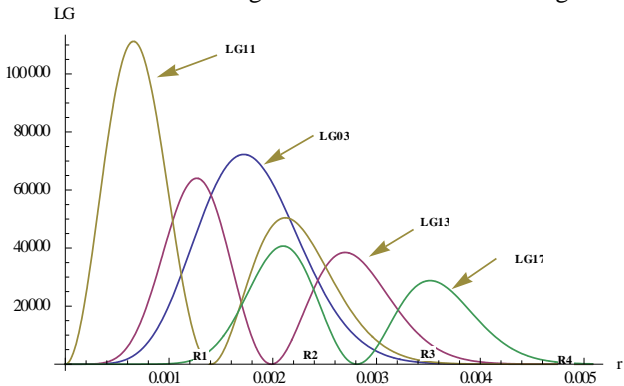

Fig. 3. Radial intensity distribution of the four OAM modes used in design of the data center receiver. R1,R2,R3 are the outer normalized radii of the annular rings of the receiver.

The simulated input beam was one of the four LG modes of Figures 1 and 3, with added white Gaussian noise. In calculating the noise, the variance was taken as proportional to the area of each detector ring. This is because thermal noise current is given as $i_{noise}^2 = \dfrac{4kT\Delta f}{R}$ where k is Boltzmann's constant, T is the temperature, R is resistance and $\Delta f$ is the range of frequency fluctuations, and we have taken Johnson's formula [22] and modeled the noise source by a current source in parallel with a resistor, using Norton's theorem [23], and thus dividing by R. Since we work at a constant frequency, and $f = \dfrac{1}{2\pi RC} \Rightarrow R = \dfrac{1}{2\pi fC}$ Capacitance is proportional to area: $C = \dfrac{\varepsilon_0 \varepsilon_r A}{d}$ where $\varepsilon_0$ is the electric constant, $\varepsilon_r$ the dielectric constant, d the separation between capacitor plates, A is plate area. So $i_{noise}^2$ ~ 1/R and R ~ 1/C ~ 1/A, therefore $i_{noise}^2$ ~ A. Therefore to obtain a noisy input vector in Matlab with appropriate noise for each ring, the random Gaussian distribution was multiplied by square root of the normalized area of each detector ring. Figure 4 shows the bit error rate as a function of the noise, averaged over input of each of the four different modes.

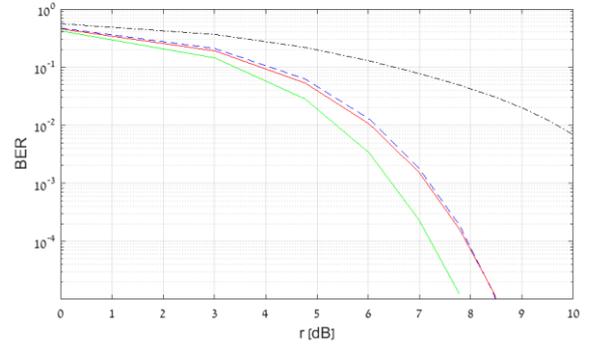

Fig. 4. Bit error rate as a function of the signal to noise per sample, averaged over input of four modes.

The bit error rate shown in the figure is the fraction of errors after the signal has been matched to the dictionary of modes using the minimal distance method of equation (4). Random Gaussian noise was added to an input mode. This was then compared to each of the dictionary modes in turn and the mode with minimal distance was registered as correct. This was repeated for $10^7$ runs. The number of correct guesses was averaged over the four modes in question since BER is different for each mode, as a result of its different intensity distribution. The x axis of the figure shows the ratio of signal to noise taken for each sample, where this was taken at the input but before implementing the minimal distance method.

The BER is unacceptable, and needs to be improved. We examined the effect of an increase in the number of rings. Figure 5 compares performance of systems of 4, 8 and 32 rings, where the BER is averaged over the four LG modes. Evidently adding more rings is advantageous but the difference between 4 and 8 rings is far more notable than that between 8 and 32 rings. This can be explained by the fact that the four modes have relatively low OAM ranging from 0 to 3, leading to a relatively small number of concentric circles of light and dark.

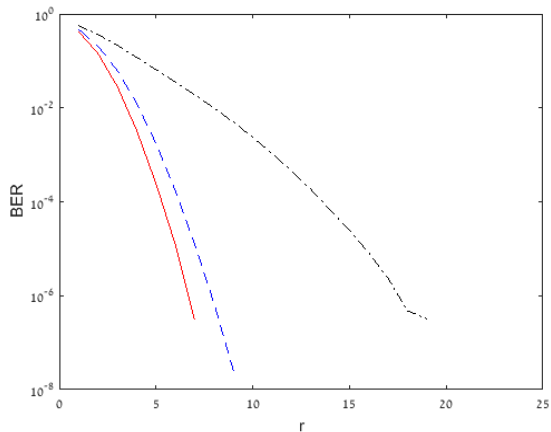

Fig 5. Bit error rate as a function of signal to noise for systems of evenly spaced rings: 4 rings (black, dot-dash line), 8 rings (blue, dashed line) and 32 rings (red, solid line).

## IV. Summary and Discussion

The next generation of communication technologies will require ever faster, flexible communication systems. Use of the orbital angular momentum of light can help meet this demand by providing a larger alphabet for encoding information, with all the flexibility, speed and resilience afforded by optical wireless communication. We have described an algorithm for decoding of an OAM system, and given a simple example for its implementation. The system is based on simple design and uses an alphabet of four modes, thus enabling a large number of bits per symbol and a significant speedup of data rate. An expansion to more than four modes is straightforward.

System performance is disappointing. This seems to be due to the small amount of data characterizing the input beam, which includes only four values. We have shown the improvement achievable by adding rings to the detector. In the controlled environment of a data center noise is not expected to pose a problem and the system presented in this paper should be sufficient, but performance needs to be further studied and improved. Future work will involve optimization of the number of detectors for minimal noise.